# IMPACTS OF STRUCTURAL FACTORS ON ENERGY CONSUMPTION IN CLUSTER-BASED WIRELESS SENSOR NETWORKS: A COMPREHENSIVE ANALYSIS

Taner Cevik[1] and Fatih Ozyurt[2]

[1]Department of Computer Engineering, Fatih University, Istanbul, Turkey
[2]Department of Software Engineering, Firat University, Elazig, Turkey

## ABSTRACT

*Limited energy is the major driving factor for research on wireless sensor networks. Clustering alleviates this energy shortage problem by reducing data traffic conveyed over the network and therefore several clustering methods are proposed in the literature. Researchers put forward their methods by making serious assumptions such as always locating single sink at one side of the topology or making clusters near to the sink with smaller sizes. However, to the best of our knowledge, there is no comprehensive research that investigates the effects of various structural alternatives on energy consumption of wireless sensor networks. In this paper, we thoroughly analyse the impact of various structural approaches such as cluster size, number of tiers in the topology, node density, position and number of sinks. Extensive simulation results are provided. The results show that the best performance about lifetime prolongation is achieved by locating a sufficient number of sinks around the network area.*

## 1. INTRODUCTION

Incredibly small sized sensor nodes have recently become available on the market with affordable prices facilitating technological improvements in microelectronics, signal processing, etc. which, in turn allowed application areas of such sensor nodes in our daily lives to expand rapidly [1-2]. These devices constitute mainly three sub-units: the processor, the sensing and the communication units [3]. Since these nodes can be self-organized without any intervention after the deployment stage, they form a Wireless Sensor Network (WSN) which is a subclass of ad-hoc networks [4]. However, there are significant differences between WSNs and their other ad-hoc counterparts. First, the nodes in traditional ad-hoc networks communicate mostly in a point-to-point manner. However, since the nodes in WSNs have limited energy sources, they prefer to communicate in a multi-hop manner. As pointed out by Akyildiz et al. [5], another important difference is that nodes in WSNs are deployed in a more intensive manner than the nodes deployed in traditional ad-hoc networks. Therefore, using the protocols and the methods utilized for ad-hoc networks will not be effective for WSNs.

As well documented in the literature, the most important drawback of these sensor nodes is the energy expenditure. Thus, in order to use thousands or millions of these devices in a topology, energy-aware protocols and architectures should be considered [6-8].

The major energy consuming unit of a sensor node is the communication unit. Raghunathan et al. [9], established that sensor nodes consume much more energy during data communication when compared with data processing. Hence, researchers have dramatically focused on developing energy-efficient communication protocols and architectures. Most prominent categories are duty-cycling methods, data-driven approaches, and clustering.

The key point in duty-cycling is defining a sub-tree of nodes in the topology that will remain awake while the others go to sleep. In this way, communication throughout the network is still active while only a portion of the nodes stay awake. Another important point is to define suitable sleep and wake-up schedules for these nodes in order to provide the sustainability of the network.

Two major subcategories constituting the data driven approach are data acquisition and data aggregation methods which aim to reduce the amount of data to be conveyed. Data acquisition is performed at signal level. In contrast, data aggregation is performed at application level. Data acquisition is adding distinct signals and transmitting data as a single aggregate. However, data aggregation is something like filtering and summarizing the original data coming from all sensor nodes.

Another very popular category of methods for lifetime prolongation is clustering. The main idea in clustering is grouping the sensor nodes depending on a number of criteria, in other words, virtually partitioning the topology into grids. Clustering can provide significant energy savings especially in high density networks. In 2011, Kumar et al. indicated that, since the plain nodes in clusters direct their data to their cluster heads, problems often encountered such as multiple routes, flooding and routing loops are eliminated or alleviated [10].

This paper presents a comprehensive analysis of the effects of the various structural factors in terms of energy consumption in WSNs. General belief about cluster-based WSNs is that in order to alleviate the hot-spot problem, clusters located near the sink should be smaller-sized than the ones further from the sink. Other possible factors that may affect the lifetime of the network are the number of tiers, the node density, the communication radio coverage radius, the number and location of the sinks. All these parameters are examined for all possible combinations in detail.

Identifying the significant role of clustering in network lifetime prolongation, the rest of the paper is organized as follows. In Section 2, we briefly describe the idea of clustering by examining a rich number of studies conducted on this topic. In Section 3, we provide information about the methods and architectures utilized during simulations. Section 4 is devoted for graphical presentation of the simulation results and discussion. Lastly, in Section 5 we provide concluding remarks.

## 2. RELATED WORK

Clustering is virtually slicing the network topology into grids (Figure 1) and grouping the sensor nodes under these grids according to a number of benchmarks.

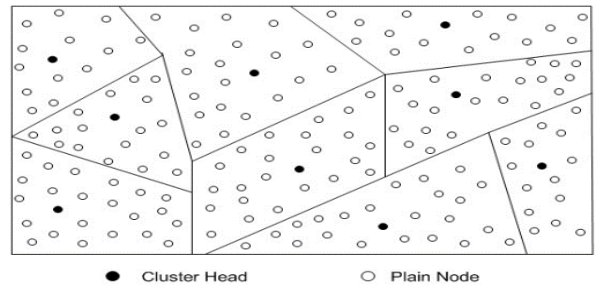

● Cluster Head    ○ Plain Node

Figure 1. Voronoi based clustering

One of the nodes in each cluster is charged with being cluster head (CH). Other nodes in the cluster which are called plain nodes gather data from the environment and deliver it to the CH node. CH is responsible for conveying the overall data gathered in its cluster to the sink. In traditional non-cluster based sensor networks, each sensor node gathers data from physical environment and aims to transmit its data to the sink somehow. If it is thought that all the nodes in the topology try to deliver their data simultaneously by flooding, a huge amount of data transmission will occur. Besides, due to the fact that all the nodes try to access the common transmission media at the same time, serious delays will occur as a result of collision prevention mechanisms. Moreover, in consequence of routing loops and multiple routes, redundant energy consumptions will result. Therefore, in terms of preventing redundant energy consumption during data transmission, clustering approach provides very significant gains by means of simplifying the communication and enhancing the scalability [11]. The objective is to find the optimum method of organizing the nodes into clusters and electing the most appropriate node as the CH in each cluster in order to achieve energy efficiency by realizing load balance among the nodes.

Many studies have been proposed about cluster-based WSNs. LEACH [12-13], is one of the first and fundamental studies conducted on WSNs and has led to many subsequent studies about clustering. The lifetime of the network is partitioned into rounds in LEACH. Cluster formation is done in an autonomous and distributed manner by the nodes without centralized supervision. Each round is divided into two phases: set-up and data transmission states. At set-up phase, each node in the topology holds a random number and depending on this number is elected to be a cluster head. Load is evenly distributed by rotating the charge of being CH among all nodes. Thus, the drainage of the nodes in the battery is delayed. Another impressive solution proposed in LEACH is the CHs making data aggregation in order to reduce the amount of data to be transmitted.

Another study of clustering following LEACH is PEGASIS [14]. Although PEGASIS is perceived as an improvement of LEACH, its basic principle is based on the chain structure rather than a cluster scheme.

HEED [15], is another work which achieves considerable improvements on energy conservation in WSNs. As in LEACH, CH selection is done periodically but not at each round. In contrast with LEACH, CH selection is not done randomly, but is rather made according to a hybrid parameter which is a combination of the residual energy levels of the nodes and a cost value called the average minimum reachability power (AMRP). AMRP is the total energy consumed by all the other nodes in the cluster if the aforementioned node becomes CH.

Clustered routing for selfish sensors (CROSS) [16] and its improved version localized game theoretical clustering algorithm (LGCA) [17] is based on the game theory for cluster formation and CH election. CROSS depends on global knowledge about the topology which is neither practical nor realistic. In contrast, LGCA employs localized information which is more suitable for energy poor WSNs.

Zhu et al. have proposed an architecture [18] in which clustering is basically performed by utilizing Hausdroff Distance [19]. The first criterion that is considered during CH selection phase is the residual energy level of the nodes. Secondly, if the residual energy levels are equal, then the proximity off the nodes is taken into account. Inter-cluster routing is performed by means of utilizing classical Bellman-Ford's shortest path approach [20].

In EECS (An Energy Efficient Clustering Scheme in Wireless Sensor Networks) [21], the residual energy levels of the nodes are again considered. Another factor impacting the CH selection is the distance between the CH candidates and the sink because inter-cluster communication is performed directly between the CHs and the sink.

In order to prevent redundant message exchange suffered during CH election phase, Cui et al., proposes an efficient idea which is called passive clustering [22]. With passive clustering, each CH candidate determines a random waiting time inversely proportional with the residual energy levels of the nodes. That is, a node with low energy level waits a longer time to announce its leadership. Therefore, the timers of the nodes with higher energy levels expire earlier and announce their leadership before the others. Thus, other nodes hearing the announcement give up the competition.

Inter-cluster communication is another challenge to be considered in cluster-based networks. Data delivered at the end of the intra-communication phase must be conveyed to the sink by the CH. This can be achieved either by single-hop or multi-hop communication. ANCAEE (A Novel Clustering Algorithm for Energy Efficiency in Wireless Sensor Networks) offers single-hop transmission for intra-cluster communication and multi-hop transmission for inter-cluster communication [23].

In addition to the studies mentioned above there have been several other studies on cluster-based sensor networks [24-28]. The next section gives details about the methods and architectures utilized in the system while analyzing the impacts of different structural variations on energy consumption.

## 3. EXPERIMENTAL ARCHITECTURE AND DETAILS

This section outlines the methodology and some significant concepts utilized in our analysis. We performed a large set of simulations with various combinations of node density, tier count, sink settlement, radio coverage, and cluster sizing. Each simulation was run until the first node death which defines the network lifetime. For performance measurement, we considered the network lifetime, since the primary challenge to be accomplished for WSNs is prolonging the working life of the network.

Instead of considering an event-based system, our simulations are based on the scenario that all nodes in the topology periodically gather data and try to transmit that data to the sink(s). For convenience, sensor nodes are assumed to be fixed and randomly distributed in a two-dimensional plane. Since all nodes potentially participate during inter-cluster communication, they do not apply any sleep-wake-up schedule.

Details about the main figures utilized during simulations are described in the following subsections.

### 3.1. Energy Consumption Model

In this paper, the classical energy model as described in LEACH is used. As is known, primary factors affecting energy consumption are the number of bits transmitted and the distance between the communicating pairs. If the distance between the communicating nodes is greater than the threshold value, then the impact of the distance on the energy consumption grows exponentially as shown in Eq. (1-3).

$$E_{snd(l,d)} = E_{snd-elec(l)} + E_{snd-amp(l,d)} \qquad (1)$$

$$E_{snd(l,d)} = \begin{cases} (l * E_{elec}) + (l * \varepsilon_{fs} * d^2) , & d<d_o \\ (l * E_{elec}) + (l * \varepsilon_{mp} * d^4) , & d \geq d_o \end{cases} \qquad (2)$$

$$E_{rcv} = l * E_{elec} \qquad (3)$$

### 3.2. Network Lifetime

Several network lifetime definitions are proposed in the literature [29-34]. Some of them consider the time in which a certain amount of the nodes die. Another idea to consider is the time after which there is a region no longer covered by the network. The one that makes the most sense and which we applied in this study is the time when the first node fails. When a node dies, it would neither be accurate nor realistic to assume that the remaining network will work well. Eventually, the node is dead and no data can be obtained from the area for which the dead node is responsible. Besides, this can result in a network partition situation which means there are two nodes which no longer can communicate with each other.

### 3.3. Cluster Head Election

Cluster Heads (CHs) have the responsibility of relaying the aggregated data of the corresponding cluster to the sink. Therefore, this heavy mission should be shared among different nodes as much as possible. Otherwise, the node assigned as CH drains its battery quickly. In this study, three types of CH election methods are analysed:

*CH Election Model 1:* Every node in a cluster runs the same algorithm similar to the one proposed in [22]. The result of the algorithm is a time value that determines the access time of a node to the common media for announcing its leadership. Other nodes hearing this announcement give up the election process and assign that node as the master node. Calculated waiting time ($Tw(i)$) at each node is reversely proportional with the distance of the node to the centre of the corresponding cluster and the residual energy level of that node:

$$Tw(i) = d(node(i), ClsCentre) / EngRes\_node(i) \qquad (4)$$

where:

d(node(i), ClsCentre) is the Euclidean distance between node(i) and the centre point of the cluster to which it belongs;

EngRes_node(i) is the residual energy of node(i).

According to Equation (4), nodes positioned around the cluster centre with higher residual energy levels wait shorter durations and therefore have a higher probability of being elected as CH than the others.

*CH Election Model 2:* This model uses a similar method to the one identified in model 1. This time, an extra parameter is involved during CH election phase as presented in Eq. (5). Nodes deployed between the centre of the corresponding cluster and the sink can be a CH. Furthermore, the distance between the node and the target sink is considered rather than the distance from the node to the centre of the cluster (Eq. (6)).

$$\text{isCHCnd} = \begin{cases} 1 \\ \infty \end{cases} \quad \text{node(i)}_{ypos} \, < \, \text{ClsCenter}_{ypos} \tag{5}$$

where:

isCHCnd is a value that determines the possibility of a node to be a CH;

node(i)ypos, and ClsCentreypos denote the (y) coordinates of node(i) and the centre point of the cluster it belongs to respectively.

Obviously, according to Eq. (6), CHs are elected among the nodes that are positioned between the centre point of the corresponding cluster and the target sink.

$$\text{Tw(i)} = (\text{isCHCnd*d(node(i),TrgSink))/EngRes\_node(i)} \tag{6}$$

where:

isCHCnd is a value that determines the possibility of a node to be a CH;

d(node(i), TrgSink) is the Euclidean distance between node(i) and the target sink;

EngRes_node(i) is the residual energy of node(i).

*CH Election Model 3:* In this method, every node in the cluster can be elected as a CH. There is no constraint like the one defined in Model 2. Again, the target is the sink(s) and nodes closer to the sink(s) with more residual energies have a greater chance to be elected as CH (Eq. (7)).

$$\text{Tw(i)} = \text{d(node(i),TrgSink)} \, / \, \text{EngRes\_node(i)} \tag{7}$$

## 3.4. Routing

Next-hop selection is performed depending on the geographical positions of the nodes. It is assumed that all nodes are aware of their relative two dimensional coordinates in the topology. Furthermore, they are assumed to be informed about the coordinates of their neighbours and the sinks settled in the topology. A number of techniques have been proposed in the literature about localization and positioning concepts. The first coming to mind is that equipping the sensor node with a Global Positioning System (GPS) receiver. However, that is not a promising solution because of deployment and cost limitations. There are other alternative solutions proposed such as lateration and angulation techniques [35]. Since it is out of scope of this study, no specific positioning method is studied in the paper.

## 3.5. Intra-Cluster Communication

Data gathered by each plain node in the cluster is delivered directly to the CH in a single-hop manner if it is in the coverage area of the sender node. Otherwise, multi-hop transmission is utilized. The packet emerging from the plain sensor node is forwarded to one of the neighbours belonging to the same cluster which is closest to the CH. The next-hop selection method for intra-cluster packet transmission is given below:

**Algorithm 1** Intra-Cluster Next-Hop Selection Method

```
findNextIntraClsHop(){
  if (isInCov(this, CH)) then
    sendPckDirectlyToCH()
  end if
  else
    distance ← ∞
    for (i←1 to numOfNgbs) do
      if (d(this,CH) < dist(ngb(i),CH))     then
        if (isInSameCls(this,ngb(i)))        then
```

```
            if (dist(ngb(i),CH) < distance) then
                distance = dist(ngb(i),CH)
                nexthop = ngb[i]
            end if
          end if
        end if
      end for
      sendIntraClsPckToNxtHop(nexthop)
    end if
}
```

## 3.6. Inter-Cluster Communication

Data aggregated at each cluster is delivered by the CHs directly to the closest sink if the sink is in the coverage area. Otherwise, the aggregated packet is forwarded to the node that is closest to the target sink. If noticed, the next hop candidate of the inter-cluster packet is not required to be in the same cluster. Next-hop calculation method for Inter-cluster communication is given below:

**Algorithm 2** Inter-Cluster Next-Hop Selection Method

```
findNextInterClsHop(){
  TrgSink ← calcTrgSink();
  if (isInCov(this, TrgSink)) then
      sendPckDirectlyToTrgSink(TrgSink)
  end if
  else
      distance ← ∞
      for (i←1 to numOfNgbs ) do
          if (d(this,TrgSink) > dist(ngb(i),TrgSink))    then
              if (dist(ngb(i),TrgSink) < distance) then
                  distance = dist(ngb(i), TrgSink)
                  nexthop = ngb[i]
              end if
          end if
      end for
      sendInterClsPckToNxtHop(nexthop)
  end if
}
```

## 3.7. Packet Structures

Length of an Intra-cluster packet is 52 bytes that takes 1625μs to transmit with the utilized radio data rate. As is known, WSNs are data-centric applications, not id-based like other traditional networks. That is, data collection centre does not deal with the ID of the data source. It is only concerned with the content. ID is only needed during forwarding operations inside the topology. Thus, there is no need to apply conventional, redundant IP or MAC addresses during in-network forwarding. It is sufficient to define short in-network unique addresses for forwarding purposes. Since all nodes are assumed to be aware of their geographical positions of themselves and their neighbours, and it is also assumed that two distinct nodes do not overlap, these relative two-dimensional coordinates constitute the ID of the nodes. Intra-cluster packet structure is presented in Figure 2.

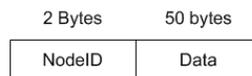

Figure 2. Intra-Cluster packet structure

CHs aggregate the data delivered by the plain nodes in its corresponding cluster and generate an Inter-Cluster packet to be transmitted to the sink. Structure of an Inter-Cluster Packet is depicted in Figure 3.

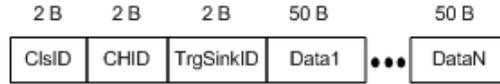

Figure 3.  Inter-Cluster packet structure

As shown in Figure 3, the first 6 octets of the Inter-Cluster packet are fixed for every Inter-Cluster packet. ClsID and CHID slots contain the IDs of the owner cluster and the corresponding CH. In this way, when the aggregated packet arrives at the data collection centre, this information identifies the region to which that packet belongs. For each round, a new sensor node is elected as a CH. Therefore, the value of the second octet part changes in each round. Field TrgSinkID denotes the ID of the target sink. Like the sensor nodes, relative two-dimensional coordinates of the sinks constitutes their IDs. Topology is virtually divided into clusters uniformly and permanently and each cluster defines its target sink at the beginning of its lifecycle, that is, the closest one relative to the centre point of the cluster. Remaining parts of the Inter-cluster packet comprises of the data delivered by each plane node in the cluster. Since the number of nodes varies for each cluster, a general formula identifying the total length of an Inter-Cluster packet is as follows:

$$LngthInterClsPck = 48 + (LngthIntraClsPck * NumOfNodes) \tag{8}$$

## 3.8. Cluster Size

One of the most important challenges encountered in WSNs is the hot-spot problem. As stated above, since sensor nodes are very tiny devices, their resources have limited capacities. The same limitation is also valid for the communication coverage radius. Sensor nodes far from the sink cannot transmit their data directly to the sink. Moreover, conveying the data directly which is actually single-hop transmission is not preferred because the energy consumed during data transmission is exponentially proportional by the distance. Therefore, multi-hop communication is preferred in WSNs. Though multi-hop communication seems to be advantageous, another vital challenge to be considered is called the hot-spot problem. Nodes closer to the sink act like a relay and convey the data incoming from the remote nodes to the sink as shown in Figure 4. Hence, all of the data traffic passes over a limited number of nodes that will cause these nodes to quickly drain the battery, which is called the hot-spot problem.

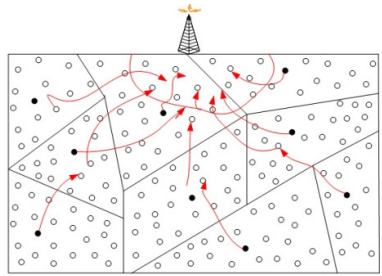

Figure 4.  Hot-spot problem

One of the promising proposals for solving the hot-spot problem is the unequal clustering method. Researchers claim that forming the cluster located closer to the sink with smaller sizes and the remote ones with larger sizes provides considerable gains in terms of energy conservation [36-42].

In this part, we inspect the effect of cluster sizes on energy consumption for three types of methods:

- Clusters closer to the sink with smaller sizes and remote clusters with larger sizes (ClsSizeModel1)

- Clusters closer to the sink with larger sizes and remote clusters with smaller sizes (ClsSizeModel2)

- All clusters with equal size (ClsSizeModel3)

### 3.9. The Number and Location of the Sinks

Generally, researchers locate a single sink at one side or at the centre of the topology. However, position and number of the sinks can affect the energy consumed in the topology. Therefore, another important point examined in this paper is the variation in network lifetime depending on the position and the number of the sink(s) in the network. Simulations were performed according to three different sink(s) localization types.

- Sink(s) located at one side of the topology (SinkPositionModel1)
- Sink located at the centre of the topology (SinkPositionModel2)
- Sink(s) positioned around the topology (SinkPositionModel3)

## 4. EXPERIMENTAL RESULTS

In this section, we analyse the performance and the impact of five structural factors on energy consumption: the number of tiers, the cluster sizes, the number and the location of the sinks, the node density and the radio coverage. While analysing the effects of these factors, different types of CH election methods are utilized and performances are compared. The simulations are performed on a 500*500 squarely shaped area where the nodes are randomly deployed. Lifetime is comprised of periodic rounds that each consists of CH election, data aggregation, intra-cluster data transmission and inter-cluster data communication phases. Topology is virtually divided into tiers and clusters.

In order to prevent common media access collisions, a MAC protocol similar to 802.11 with RTS/CTS mechanism is used. As mentioned in the previous section, we employed the classical energy calculation model that depends primarily on the distance. Parameters utilized during simulations are given in Table1.

Table 1. Simulation parameters

| | |
|---|---|
| Radio transmission data rate | 250 Kbps |
| $d_0$ (threshold distance) | 85 m |
| $R_0$ (coverage radius) | 100 m |
| Eelec | 50 nJ/bit |
| $\varepsilon_{fs}$ | 10 pJ/bit/m$^2$ |
| $E_{mp}$ | 0.0013 |

As mentioned earlier, the impacts of cluster sizing, the number and the location of the sink(s), the number of tiers and the node density on energy consumption are represented comprehensively. Simulations are performed under three structural titles depending on how the clusters are sized. As clarified previously, these are: clusters closer to the sink with smaller sizes and remote clusters with larger sizes (ClsSizeModel1), clusters closer to the sink with larger sizes and remote clusters with smaller sizes (ClsSizeModel2) and all clusters with equal sizes (ClsSizeModel3).

### 4.1. Clusters Close to the Sink with Smaller Sizes (ClsSizeModel1)

Another alternative solution proposed by the researchers is to slice the topology into tiers. As shown in Figure 5, incrementing the number of tiers redundantly makes a negative impact on the lifetime of the network. With a topology of containing 500 nodes, the best performance is provided by utilizing 2 tiers. SelectCH_Centre, SelectCH_EN_AfterCentre and SelectCH_EN_withNoCons correspond to the CH election schemes CH Election Model 1, CH Election Model 2 and CH Election Model 3 respectively.

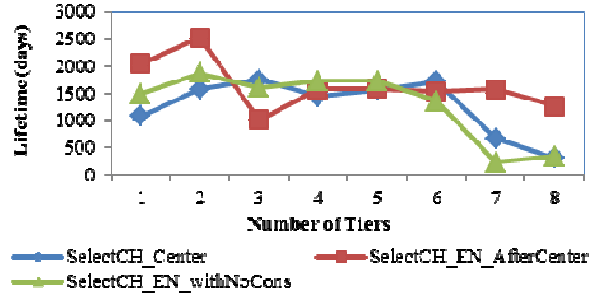

Figure 5. Change in the network lifetime depending on the number of tiers

Another possible factor that may affect the network lifetime is the number of nodes in the network that is node density. By holding the size of the topology area constant, increasing the number of nodes increases the node density. However, more nodes means more data packets to be transmitted. Thus, the load is observed to be shared by the time the density increases; however, increase in the network traffic balances this factor. Figures 6-8 show the lifetime performance depending on the node density for the networks that the sink(s) positioned one side, sink at the centre and sink(s) surrounding the nodes respectively.

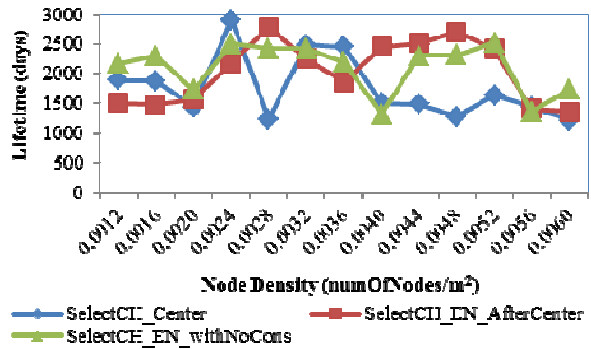

Figure 6. Change in the network lifetime depending on the node density for (SinkPositionModel1, ClsSizeModel1) pair

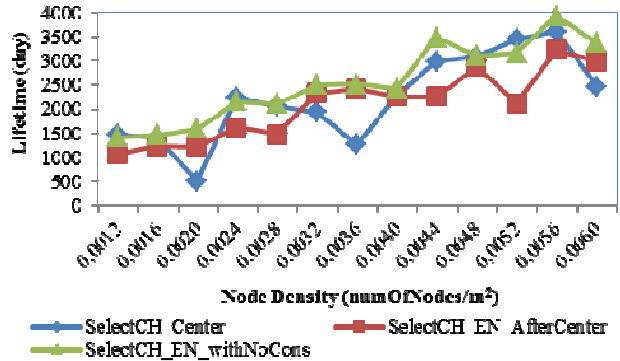

Figure 7. Change in the network lifetime depending on the node density for (SinkPositionModel2, ClsSizeModel1) pair

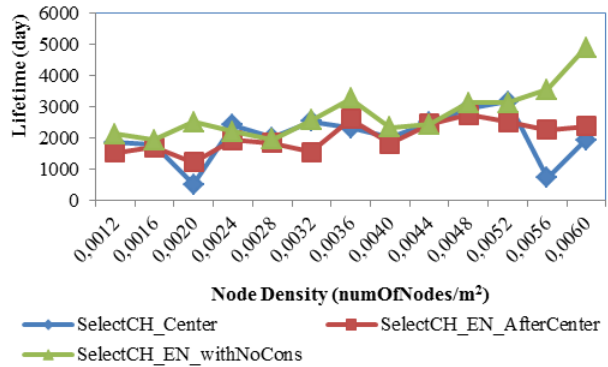

Figure 8. Change in the network lifetime depending on the node density for (SinkPositionModel3, ClsSizeModel1) pair

Figures 6-8 clarify that a topology model surrounded with the sinks provides the best performance in terms of network lifetime.

Another possible factor that can affect the network performance is the number of sinks in the system. We examined whether an increase in the number of sinks prolongs the lifetime of the network.

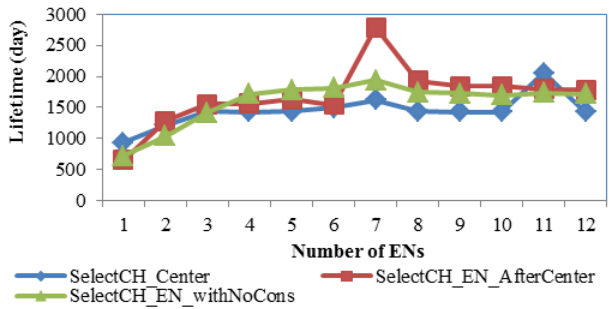

Figure 9. Change in the network lifetime depending on the number of sinks for (SinkPositionModel1, ClsSizeModel1) pair

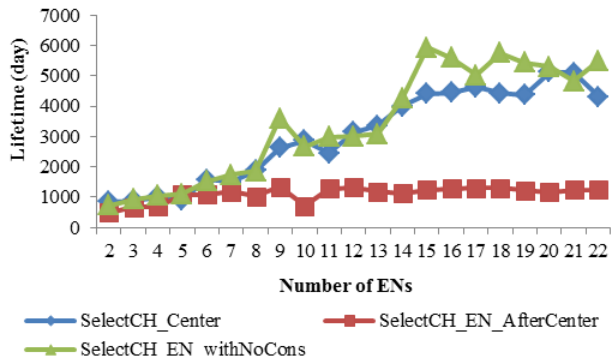

Figure 10. Change in the network lifetime depending on the number of sinks for (SinkPositionModel3, ClsSizeModel1) pair

Figures 9-10 reprove that with SinkPositionModel3, network lifetime is doubled by increasing the number of sinks in the network. Since, the sensor nodes have limited coverage capacities and the energy consumption is proportional with the communication distance, multi-hop communication is preferred in WSNs. In our simulation model, CHs directly forward the aggregated data to the sink(s) if they are in the communication range. However, this causes an extra burden. As mentioned in the previous section, energy consumption of the communication unit is exponentially proportional with the distance between the receiver and the sender. This idea is supported by Figures 11-13. Radio coverage improvement does not bring an extra advantage in terms of energy conservation.

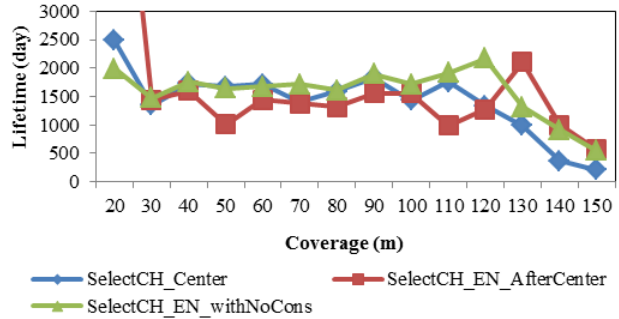

Figure 11. Change in the network lifetime depending on the radio coverage for (SinkPositionModel1, ClsSizeModel1) pair

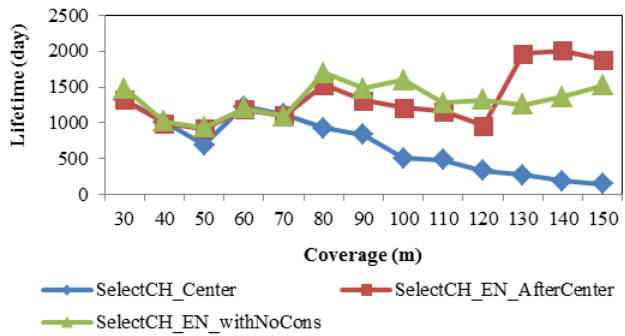

Figure 12. Change in the network lifetime depending on the radio coverage for (SinkPositionModel2, ClsSizeModel1) pair

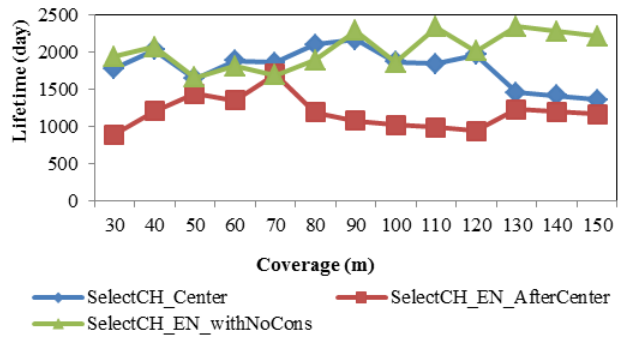

Figure 13. Change in the network lifetime depending on the radio coverage for (SinkPositionModel3, ClsSizeModel1) pair

## 4.2. Clusters Close to the Sink with Larger Sizes (ClsSizeModel2)

Figures 14-21 present the changes occur in the network lifetime depending on the parameters presented in the previous section. This time, clusters closer to the sink(s) have larger sizes and the further ones with smaller sizes.

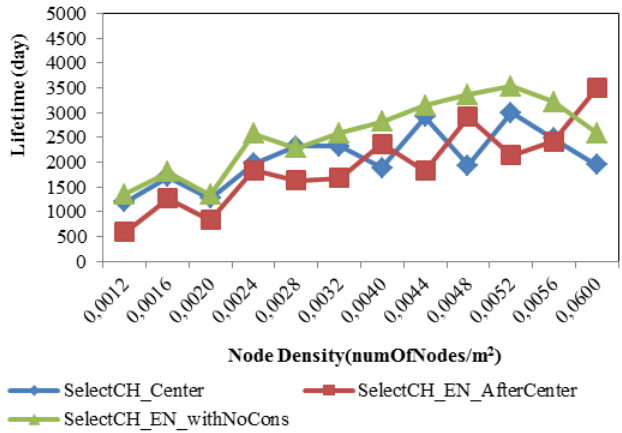

Figure 14. Change in the network lifetime depending on the node density for (SinkPositionModel1, ClsSizeModel2) pair

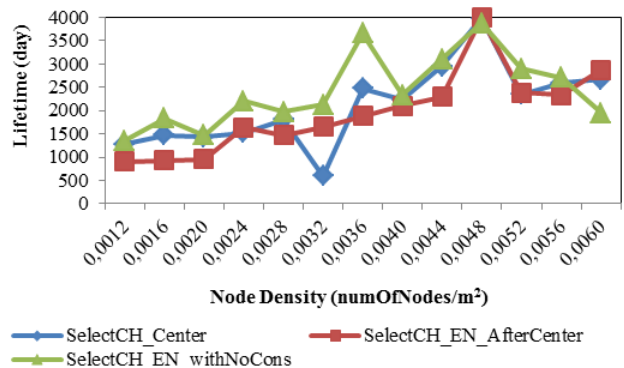

Figure 15. Change in the network lifetime depending on the node density for    (SinkPositionModel2, ClsSizeModel2) pair

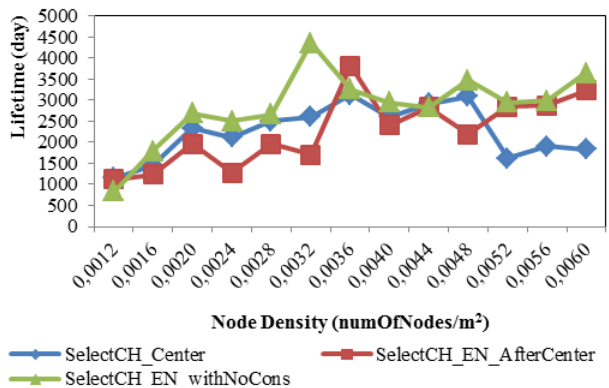

Figure 16. Change in the network lifetime depending on the node density for (SinkPositionModel3, ClsSizeModel2) pair

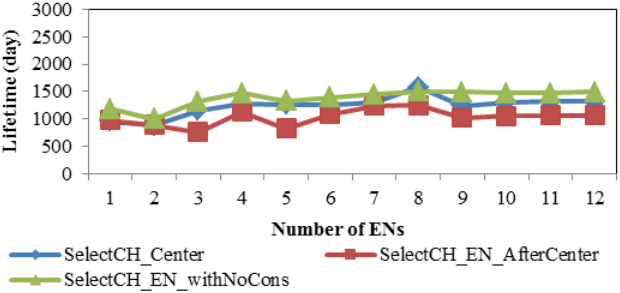

Figure 17. Change in the network lifetime depending on the number of sinks for (SinkPositionModel1, ClsSizeModel2) pair

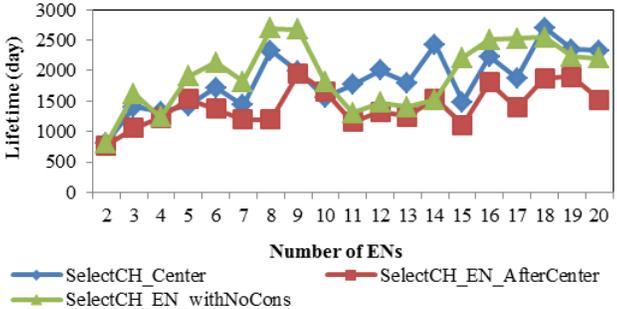

Figure 18. Change in the network lifetime depending on the number of sinks for (SinkPositionModel3, ClsSizeModel2) pair

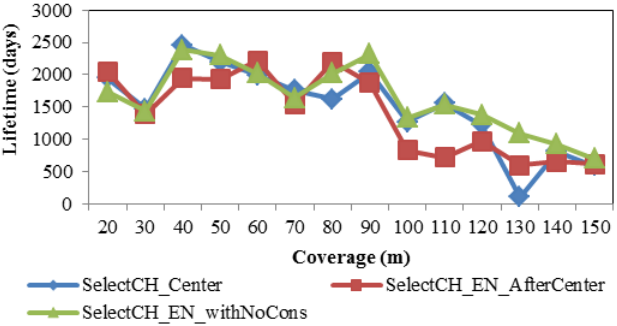

Figure 19. Change in the network lifetime depending on the radio coverage for (SinkPositionModel1, ClsSizeModel2) pair

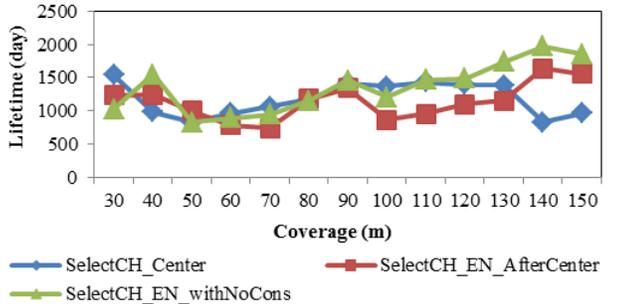

Figure 20. Change in the network lifetime depending on the radio coverage for (SinkPositionModel2, ClsSizeModel2) pair

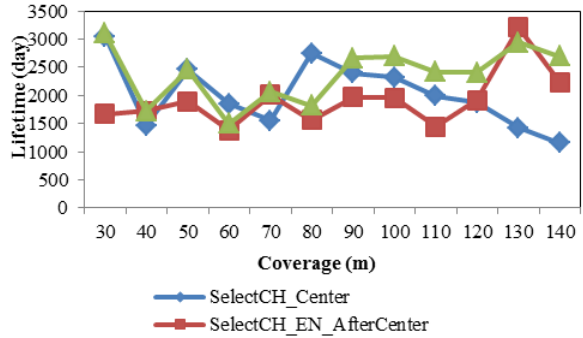

Figure 21. Change in the network lifetime depending on the radio coverage for (SinkPositionModel3, ClsSizeModel2) pair

### 4.3. All Clusters with Equal Sizes (ClsSizeModel3)

Finally, Figures 22-29 present the changes that occur in the network lifetime depending on the parameters presented in the previous section. This time, all the clusters in the network have equal sizes.

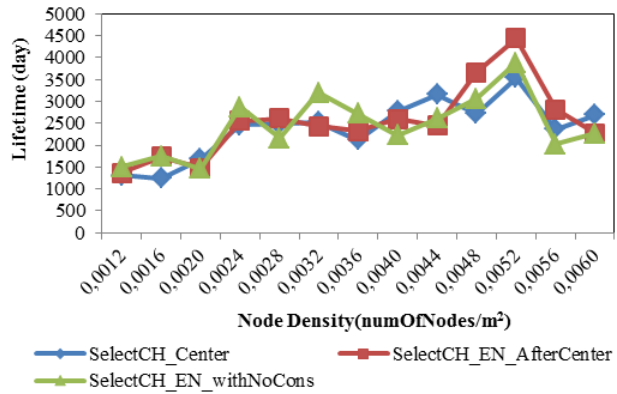

Figure 22. Change in the network lifetime depending on the node density for (SinkPositionModel1, ClsSizeModel3) pair

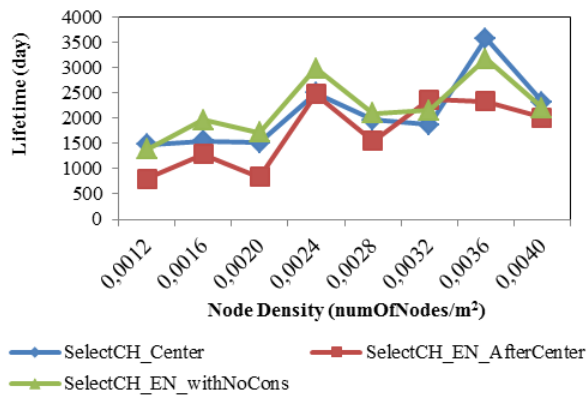

Figure 23. Change in the network lifetime depending on the node density for (SinkPositionModel2, ClsSizeModel3) pair

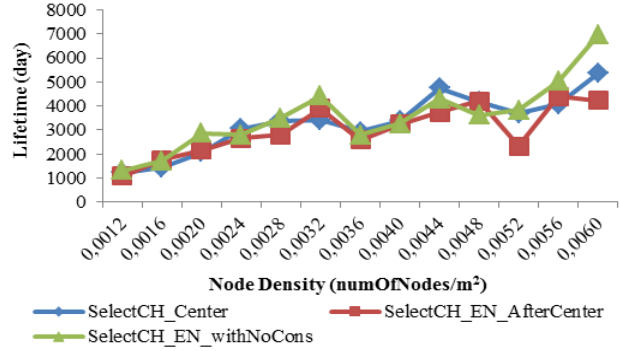

Figure 24. Change in the network lifetime depending on the node density for (SinkPositionModel3, ClsSizeModel3) pair

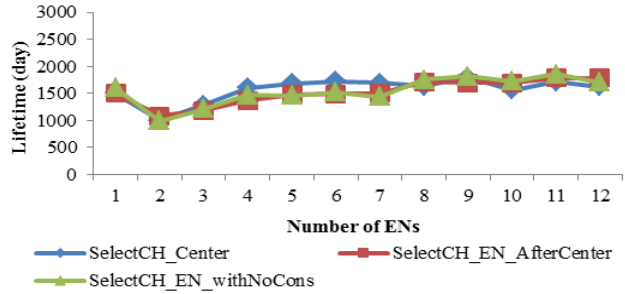

Figure 25. Change in the network lifetime depending on the number of sinks for (SinkPositionModel1, ClsSizeModel3) pair

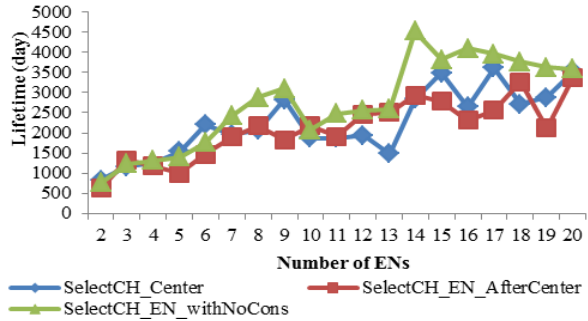

Figure 26. Change in the network lifetime depending on the number of sinks for (SinkPositionModel3, ClsSizeModel3) pair

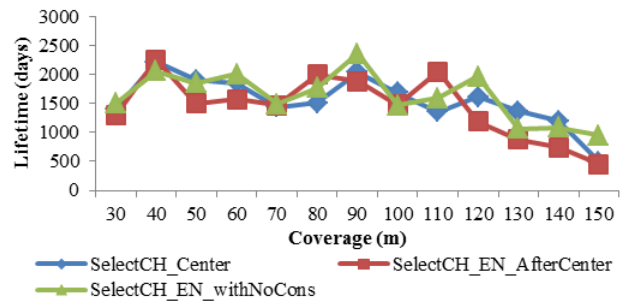

Figure 27. Change in the network lifetime depending on the radio coverage for (SinkPositionModel1, ClsSizeModel3) pair

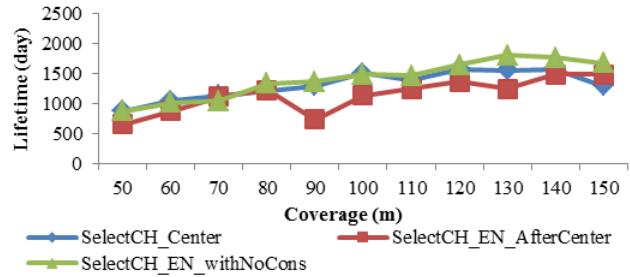

Figure 28. Change in the network lifetime depending on the radio coverage for (SinkPositionModel2, ClsSizeModel3) pair

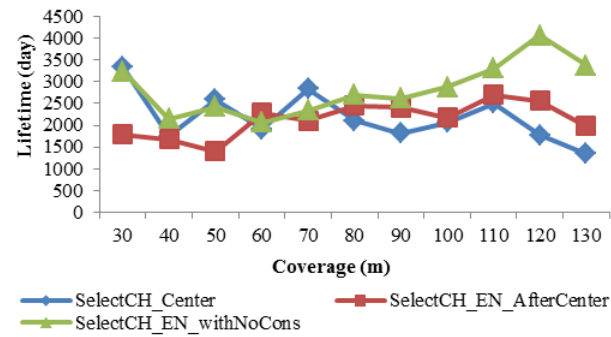

Figure 29. Change in the network lifetime depending on the radio coverage for (SinkPositionModel3, ClsSizeModel3) pair

## 5. CONCLUSIONS

The primary energy-consuming unit of a sensor node is the communication unit. It is crucial that while designing protocols, methods, and architectures for WSNs, this energy constraint problem should be considered. Up to now, several research activities performed and various methods have been proposed about cluster-based WSNs. This paper presents a brief analysis of the effects of the various structural factors in terms of energy consumption in cluster-based WSNs. General belief about cluster-based WSNs is that in order to alleviate the hot-spot problem, clusters located near the sink should be smaller-sized than the remote ones. Other possible factors that may affect the lifetime of the network are the number of tiers, the node density, the communication radio coverage radius, the number and location of the sinks. All these parameters are examined for all possible combinations in detail. Depending on our simulations, the best performance in terms of the network lifetime is provided by positioning the sinks around the network. Increasing the node density up to a level is another factor that affects the energy consumption positively. Also, it is proved that sizing the clusters closer to the sink smaller than those further away enhances the energy conservation. Furthermore, it is also clarified that larger radio coverage does not have a definite positive effect in terms of energy conservation.

## AUTHORS

Taner Cevik received the B. S., M.S. and Ph.D. degrees in computer engineering from Istanbul Technical University in 2001, Fatih University in 2008, and Istanbul University in 2012 respectively. In 2006, he joined the Department of Computer Engineering, Fatih University, as a research assistant, and in 2010 became an instructor at the same university. Since 2013, he has served as an assistant professor at Fatih University.

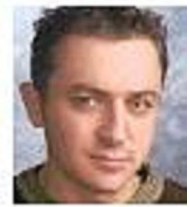

Fatih Ozyurt received the B.S. and M.S. degrees in computer engineering from Eastern Mediterranean University Cyprus, and Fatih University Istanbul Institute of Science in 2011 and 2014, respectively. He continues his education as a PHD student in Software Engineering at Firat University in Elazig.

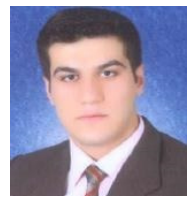